# Resolving Kane's Puzzle in Oblique Collisions of Rigid Bodies

Xueqiang Wang[a,†], Qi Su[b,†], Siping Li[b,*]

[a] *National Engineering Research Center of Light Alloy Net Forming and State Key Laboratory of Metal Matrix Composites, School of Materials Science and Engineering, Shanghai Jiao Tong University, Dongchuan Road 800, Shanghai 200240, China*

[b] *Department of Engineering Mechanics, School of Ocean and Civil Engineering, Shanghai Jiao Tong University, Dongchuan Road 800, Shanghai 200240, China*

[†] These authors contributed equally to this work.
[*] Corresponding author.
E-mail address: lisp@sjtu.edu.cn (Siping Li)

**Abstract**

We examined the asymmetric deformation in collisions and the transition conditions from oblique to normal collisions and non-collisions to address the problem of oblique collisions of rigid bodies in classical mechanics. A closed solution satisfying the fundamental equations and adhering to the energy conservation law without introducing new material parameters was derived. The solution exhibited a nonlinear relationship between post-collision velocity and initial state parameters, contrasting with the linear results of existing studies. This solution avoided the fallacy in Whittaker's hypothesis that kinetic energy might increase after a collision. Consequently, the solution presented herein fundamentally resolves Kane's puzzle, previously overlooked in classical mechanics.

# 1 Introduction

Oblique collisions are ubiquitous nonlinear mechanical phenomena that involve contact and catastrophe [1–6]. These events range from the mundane, such as an object's bounce after falling [7–9] and hard objects hitting a helmet [10,11], to accidents, like vehicle crashes [12–14] and meteorite descents [15,16]. At the microscopic level, particle collision [17–20] showcases the basic laws of oblique collisions, while at the macroscopic scale, bullets piercing steel plates [21–23] structural components absorb impact energy [24,25], and feet rubbing against the ground during locomotion [26,27] exhibit the universalities of these interactions. In contemporary sports, various ball games consistently demonstrate the diverse nature of oblique collisions [28–34].

Classical mechanics, governed by Newton's laws, represents the first axiomatized branch of physics. Augmenting Coulomb's law of friction and Newton's theory of collision restitution is necessary to derive realistic and unique solutions to material-related problems of friction and collision within classical mechanics. However, the coupled problem of friction and collision remains unsolved. This deficiency in classical mechanics has drawn considerable attention from researchers.

Initially, Whittaker (1904) [35,36] proposed the hypothesis that the tangential impulse equals the normal impulse multiplied by the Coulomb friction coefficient. The oblique collision theory has gained widespread acceptance and is readily applied in engineering. However, 80 years later, Kane (1984) [37,38] identified that Whittaker's hypothesis can sometimes lead to a paradox where energy increases after a collision, violating the law of energy conservation, indicating a fundamental flaw in Whittaker's hypothesis.

Kane's puzzle has garnered significant attention from researchers. Keller (1986) [39], Smith (1991) [40], Brach [41], Chatterjee [42], and Stronge [43,44] have attempted various modifications to Whittaker's hypothesis. However, none have proven successful. Additionally, Rémond [45], Donahue [46], Kuninaka [47], Louge [48], Saitoh [49], and Djerassi [50] have explored a variety of material collision models in their studies. Although progress has been made in their respective areas, the essence of oblique collisions involving rigid-body models has not been thoroughly explored.

Whittaker's hypothesis exhibits discontinuity at the critical transition point from a oblique to a normal collision. The assumed tangential impulse $I_\tau = \mu I_n$ inevitably generates a tangential velocity at the contact point, which does not originate from normal collisions, potentially leading to spurious or erroneous estimations of the kinetic energy of the system. Unlike the symmetric deformation characteristic of normal collisions, oblique collisions display asymmetric deformation features (Fig. 1) [51–54]. Therefore, the tangential impulse in oblique collisions should encompass both oblique and tangential deformation impulses. This inclusion considered the tangential deformation and its recovery impulse, potentially resulting in the reversal of tangential velocity at the contact point.

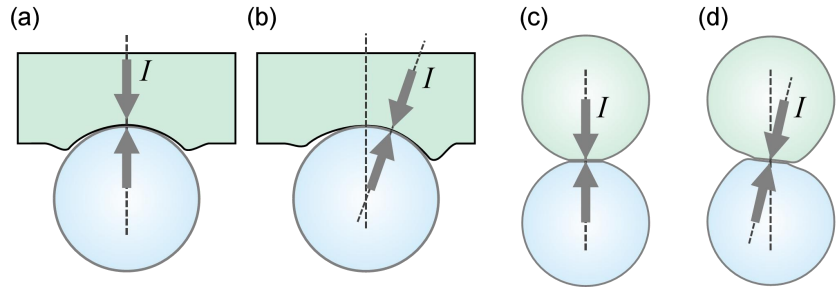

**Fig. 1.** Schematic of collision deformation states. (a) (c) Normal collision and (b) (d) Oblique collision.

The academic community has been perplexed for over a century by the issue of rigid-body friction collisions, largely due to the lack of in-depth analysis and correct understanding of the features and mechanisms of oblique collisions. Consequently, previous research on oblique collisions has lacked a proper theoretical foundation [55–59].

Despite extensive research, Kane's puzzle remains an outstanding problem that continues to captivate and divide the scientific community. This study aims to offer a novel perspective on resolving Kane's puzzle in oblique collisions of rigid bodies. The preliminary sections of this paper have discussed the flaws of Whittaker's hypothesis and the mechanics of oblique collisions. In Section 2, we present a methodology to resolve Kane's puzzle without introducing new material parameters, providing a coherent solution to the puzzle. In Section 3, we analyze and discuss the proposed solution with models from other researchers. Finally, the main conclusions and implications of the proposed solution are summarized in Section 4.

## 2 Solution methodology

### 2.1 Oblique collision model

This work presents a simplified model for analyzing the oblique collision of a homogeneous sphere impacting a fixed plane. The problem can be reduced to a planar collision in the osculating plane because the normal rotation during the collision is inherently balanced, as depicted in Fig. 2.

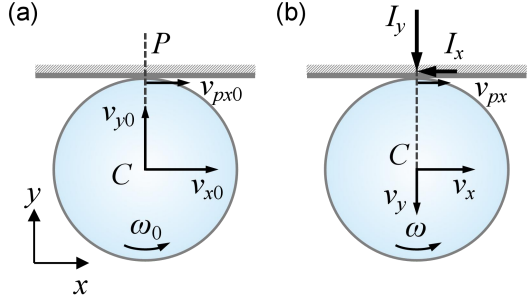

**Fig. 2.** Schematic of a friction collision involving a sphere of mass *m* and radius *r* impacting a fixed plane. (a) Pre-collision: the center-of-mass and angular velocities are $v_{x0}$, $v_{y0}$ ($\geq 0$), and $\omega_0$, respectively, and the tangential velocity at the contact point is $v_{px0} = v_{x0} - \omega_0 r$ ($\geq 0$). (b) Post-collision: the corresponding variables are $v_x$, $v_y$, $\omega$ ($\leq 0$), $v_{px} = v_x - \omega r$.

$I_x$ and $I_y$ in Fig. 2(b) represent the tangential and normal forces on the sphere, respectively. The collision dynamics equations are

$$m(v_y - v_{y0}) = I_y, \tag{1}$$

$$m(v_x - v_{x0}) = -I_x, \tag{2}$$

$$J_c(\omega - \omega_0) = rI_x. \tag{3}$$

In the normal direction, Newton's theory of collision recovery applies

$$v_y = -ev_{y0}, \tag{4}$$

where *e* denotes the normal restitution coefficient.

Introducing dimensionless parameters

$$k = \frac{v_{px}}{v_{px0}}, \tag{5}$$

$$i_x = \frac{I_x}{\mu I_y}, \tag{6}$$

$$\alpha = \frac{1}{2} \frac{(J_c + mr^2)\mu(1+e)v_{y0}}{J_c v_{px0}} = \frac{j_p \mu I_y}{2m v_{px0}}, \tag{7}$$

where $\mu$ represents the kinematic friction coefficient, $J_c$ corresponds to the moment of inertia about the center of mass, and $j_p$ denotes the dimensionless moment of inertia around the contact point,

$$j_p = \frac{J_p}{J_c} = \frac{J_c + mr^2}{J_c}. \tag{8}$$

Substituting Eqs. (5)–(7) into Eqs. (1)–(4) and simplifying results in

$$i_x = \frac{1}{2\alpha}(1-k), \tag{9}$$

where $i_x$ and $k$ correspond to functions of the initial state parameter $\alpha$, with $0 < \alpha \leq +\infty$. Note that the friction coefficient is included in $\alpha$. Additional definite solution conditions must be imposed to obtain a unique solution for oblique collisions.

Equation (9) must satisfy the critical connection conditions from oblique collisions to normal collisions ($\alpha \to +\infty$) and from oblique collisions to non-collisions ($\alpha \to 0$) to ensure continuity of collision parameter variations.

**2.2 Connection conditions of normal collisions**

When the normal velocity significantly exceeds the tangential velocity, indicating the dominance of normal collisions, the tangential velocity at the contact point, calculated using the friction law $I_x = \mu I_y$, must be reversed. Consequently, $v_{px0} \to 0^+$ and $v_{px} \to 0^-$ occur at the critical point of transition from oblique to normal collisions, i.e.

$$k = \lim_{\alpha \to +\infty} \frac{v_{px}}{v_{px0}} = -1. \tag{10}$$

Additionally, the following conditions must be satisfied for normal collisions

$$i_x = \lim_{\alpha \to +\infty} \frac{I_x}{\mu I_y} = 0. \tag{11}$$

**2.3 Connection conditions of non-collisions**

When the normal velocity is small compared to the tangential velocity, approaching zero, the effect of normal collision diminishes, the collision deformation is minimal, and the friction component predominates. Therefore, the conditions $v_{px} \to v_{px0}$, $I_x \to \mu I_y$ apply. Therefore, the following conditions are available at the critical point of non-collisions

$$k = \lim_{\alpha \to 0} \frac{v_{px}}{v_{px0}} = 1, \tag{12}$$

$$i_x = \lim_{\alpha \to 0} \frac{I_x}{\mu I_y} = 1. \tag{13}$$

**2.4 Solution**

The above analysis shows that a tangential impulse arises from the combined effects of friction and tangential deformation. Directly describing tangential deformation is challenging; an alternative approach involves representing deformation as a change in tangential velocity (or the combination of tangential velocities before and after the collision). This concept, inspired by Eq. (9), ensures dimensional consistency. Therefore, the tangential impulse can be assumed to be a linear combination of pre and post-collision tangential velocities

$$I_x = A\mu I_y + Bmv_{px0} + Cmv_{px},$$

where $A$, $B$, and $C$ are constants; $I_y = m(1+e)v_{y0}$ and $v_{px0}$ represent the initial condition values of the collision. The above assumption can be equivalently simplified as

$$i_x = \frac{I_x}{\mu I_y} = a + bk, \tag{14}$$

where $a$ and $b$ are constants to be determined. We find $a = b = 1/2$ by substituting the connection conditions in Eqs. (10)–(13) into Eq. (14). Therefore, Eq. (14) becomes

$$i_x = \frac{1}{2}(k+1). \tag{15}$$

Solving Eq. (9) and (15) simultaneously yields

$$k = \frac{1-\alpha}{1+\alpha}. \tag{16}$$

Hence, we obtain

$$i_x = \frac{1}{1+\alpha}. \tag{17}$$

Equations (16) and (17) represent the solutions for oblique collisions under the rigid body model, where $-1 < k \leq 1$. This solution avoids the inherent defects of Whittaker's hypothesis because it satisfies the connection conditions of both normal and non-collisions.

According to the kinetic energy theorem for collisions, the change in the kinetic energy of the system during a collision is equivalent to the work done by the impulse. This allowed us to calculate the work done by multiplying the impulse by the average velocity at the contact point. Furthermore, we specifically investigated the energy changes in the system excluding the normal kinetic energy, i.e., the change in kinetic energy in the tangential and rotational directions given by

$$\Delta E_{\tau\omega} = -\frac{1}{2}I_x(v_{px} + v_{px0}). \tag{18}$$

The corresponding dimensionless energy is

$$\Delta e_{\tau\omega} = \frac{\Delta E_{\tau\omega}}{\mu I_y v_{px0}} = -i_x \cdot \frac{1}{2}(1+k) = -\frac{1}{4\alpha}(1-k^2) \leq 0, \tag{19}$$

where $-1 < k \leq 1$. Equation (19) demonstrates that the solution presented in this work adheres to the fundamental law that energy loss occurs during a collision. According to Newton's theory of collision recovery (Eq. (4)) and the fact that $0 < e < 1$, it is clear that the velocity in the normal direction decays. Therefore, the energy conservation law is satisfied in the normal, tangential, and rotational directions, avoiding Kane's puzzle.

## 3  Results and discussion

The tangential impulse in Whittaker's hypothesis is

$$I_x = \mu I_y. \tag{20}$$

Substituting Eq. (20) into Eq. (5) gives the dimensionless tangential impulse as

$$i_{xW} = 1. \tag{21}$$

Substituting Eq. (21) into Eq. (9) (and Eq. (19) )gives the dimensionless tangential velocity (and energy) in Whittaker's hypothesis as (converted to the parameters used in this work)

$$k_W = -2\alpha + 1. \tag{22}$$

and

$$\Delta e_{\tau\omega} = -\frac{1}{2}(1+k) = \alpha - 1. \tag{23}$$

When $\alpha > 1$, $\Delta e_{\tau\omega} > 0$, leading to the paradox of energy non-conservation.

Building on this, Smith [40] introduced a correction to the tangential impulse using a coefficient $\eta$, i.e. $I_x = \eta\mu I_y$. Hence, Smith's solutions are

$$k_S = -2\eta\alpha + 1, \tag{24}$$

$$i_{xW} = 1, \tag{25}$$

where $\eta$ can be assigned a value of 0.4.

Keller [39] bifurcated the collision process using the instant at which the tangential sliding velocity becomes zero, applying Whittaker's hypothesis for each phase before and after the direction of sliding changes. The normal impulse $I_y$ is accordingly divided into $I_{y1}$ and $I_{y2}$, with one scenario being $I_{y1} = 2I_y/3$ and $I_{y2} = I_y/3$, translating to $I_{x1} = 2\mu I_y/3$ and $I_{x2} = \mu I_y/3$ respectively. This approach proves that one of Keller's solutions is represented as follows

$$k_K = \begin{cases} -\dfrac{4}{3}\alpha + 1 & (k > 0) \\ -\dfrac{2}{3}\alpha + 1/2 & (k < 0) \end{cases} \tag{26}$$

$$i_K = \begin{cases} \dfrac{2}{3} & (k > 0) \\ \dfrac{1}{3} & (k < 0) \end{cases} \tag{27}$$

The solutions to the tangential velocity presented in this work, along with those of Whittaker [35],[36], Smith [40], and Keller [39] (all converted to the parameters used herein, respectively) are depicted in Fig. 3. The results from Whittaker as well as the corrections by Smith and Keller, exhibit a linear variation, which inevitably leads to the failure zone where energy is not conserved. The tangential impulse in this work varies nonlinearly, converging to -1 as *α* increases, but not below -1, thus avoiding Kane's puzzle.

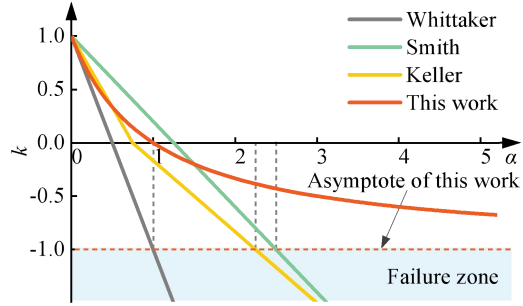

**Fig. 3.** Variation curve of dimensionless tangential velocity *k* with collision parameter *α*. The zone with *k* < -1 is the failure zone, meaning that the energy will not be conserved.

Figure 4 depicts the relationship between tangential impulse $i_x$ and parameter *α*. The $i_x$ values in this work exhibit a nonlinear variation, showing consistency with Whittaker's results when *α* equals zero. In contrast, solutions proposed by other scholars remain constant, leading to an overestimation of the tangential impulse magnitude to varying degrees. Consequently, employing their methods results in the calculation of increased kinetic energy after a collision, which violates the actual physical laws. This discrepancy underscores the importance of considering the nonlinear nature of tangential impulse in accurately modeling and predicting physical phenomena.

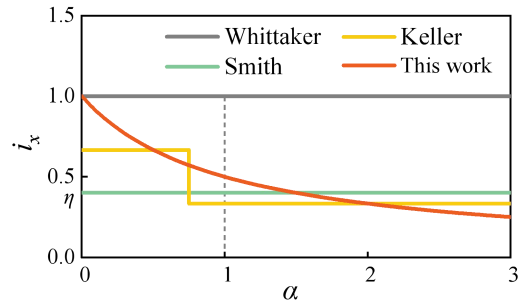

**Fig. 4.** Variation curve of dimensionless tangential impulse $i_x$ with collision parameter *α*.

The tangential impulse ($I_x$) in this work, derived from Eq. (17), is

$$I_x = \frac{\mu}{1+\alpha} I_y, \qquad (28)$$

This indicates that the friction coefficient ($\mu'$) of the solutions presented in this work is

$$\mu' = \frac{\mu}{1+\alpha}. \qquad (29)$$

The relationship between $\mu'$ and *α* is depicted in Fig. 5. Notably, $\mu'$ exhibits a nonlinear decrement with increasing $α$, contradicting Whittaker's hypothesis of it being a constant. As a normal collision ($α → +∞$) occurs, $μ'$ converges to zero. Conversely, in the non-collision ($α → 0$), $μ'$ equals $μ$. This observation suggests that Whittaker's hypothesis holds exclusively under sliding friction conditions.

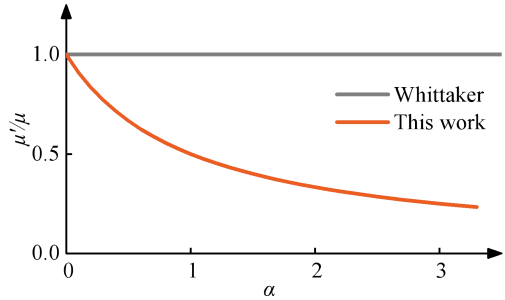

**Fig. 5.** Variation curve of dimensionless friction coefficient with collision parameter $α$.

## 4 Conclusions

In summary, the linear solutions derived by previous researchers based on assuming initial parameters did not effectively resolve Kane's puzzle. This work presents a nonlinear closed-form solution that considers the coupling effect of friction and tangential deformation without introducing new material parameters. The proposed mechanism addresses the limitation that collision deformation is generally overlooked in traditional rigid-body models. This solution eliminates the possibility of an increase in the kinetic energy of the system after a collision, providing the first effective solution to the oblique collision problem in a rigid-body model. The research methodology presented in this study can be readily applied to solve oblique collision problems in more complex models.

**Conflict of Interest**

The authors declare that they have no conflict of interest.

**Acknowledgments**

This work is supported by the National Natural Sciences Foundation of China (grant number 51878407).

**Author Contributions**

**Xueqiang Wang**: Conceptualization, Methodology, Validation, Writing - original draft, Writing - review & editing, Visualization

**Qi Su**: Conceptualization, Methodology, Validation, Writing - original draft, Writing - review &

editing, Visualization

**Siping Li**: Investigation, Resources, Supervision, Funding acquisition.